\documentclass[twocolumn]{pasj02}

\jyear{2026}
\Received{2026/02/13}
\Accepted{2026/03/04}
\Published{$\langle$publication date$\rangle$}
\usepackage[switch,mathlines]{lineno}

\usepackage{url}
\usepackage{comment}
\usepackage{multirow}
\usepackage{graphicx}	
\usepackage{amsmath}	
\usepackage{natbib}

\def\kms{km s$^{-1}$}
\renewcommand{\thefootnote}{\fnsymbol{footnote}}

\everymath{\displaystyle}

\begin{document}

\title{The trigonometric parallax of IRAS 23385+6053 and physical properties of molecular clouds based on the VLBI astrometry}

\author{Shota \textsc{Hamada}\altaffilmark{1}}
\author{Mikito \textsc{Kohno}\altaffilmark{2,3}\altemailmark\orcid{0000-0003-1487-5417}\email{kohno.ncsmp@gmail.com}}
\author{Toshihiro \textsc{Omodaka}\altaffilmark{1}\altemailmark\email{omodaka.toshihiro@gmail.com}}
\author{Nobuyuki  \textsc{Sakai}\altaffilmark{4, 5}\altemailmark\orcid{0000-0002-5814-0554}\email{nobuyuki@narit.or.th}}
\author{Riku \textsc{Urago}\altaffilmark{6,7}\orcid{0000-0002-6356-4255}}
\author{Takumi \textsc{Nagayama}\altaffilmark{5}}
\author{Hideyuki \textsc{Kobayashi}\altaffilmark{4}\orcid{0000-0001-8066-1631}}
\author{Yuji \textsc{Ueno}\altaffilmark{5}}

\altaffiltext{1}{Graduate School of Science and Engineering, Kagoshima University, 1-21-35 Korimoto, Kagoshima, Kagoshima 890-0065, Japan}
\altaffiltext{2}{Curatorial division, Nagoya City Science Museum, 2-17-1 Sakae, Naka-ku, Nagoya, Aichi 460-0008, Japan}
\altaffiltext{3}{Department of Physics, Graduate School of Science, Nagoya University, Furo-cho, Chikusa-ku, Nagoya, Aichi 464-8602, Japan}
\altaffiltext{4}{National Astronomical Research Institute of Thailand (Public Organization), 260 Moo 4, T. Donkaew, A. Maerim, Chiang Mai, 50180, Thailand}
\altaffiltext{5}{Mizusawa VLBI Observatory, National Astronomical Observatory of Japan, 2-12 Hoshigaoka-cho, Mizusawa, Oshu, Iwate 023-0861, Japan}
\altaffiltext{6}{Astrobiology Center, 2-21-1 Osawa, Mitaka, Tokyo 181-8588, Japan}
\altaffiltext{7}{National Astronomical Observatory of Japan (NAOJ), National Institutes of Natural Sciences (NINS), 2-21-1 Osawa, Mitaka, Tokyo 181-8588, Japan}

\maketitle

\begin{abstract}
We performed very long baseline interferometry (VLBI) observations to measure the trigonometric parallax of H$_2$O maser sources in the outer massive star-forming region IRAS 23385+6053 using the VLBI Exploration of Radio Astrometry (VERA) in Japan.
{The annual parallax is $\pi=0.460 \pm 0.086$~mas, which corresponds to a distance of $2.17^{+0.50}_{-0.34}$ kpc, roughly half the kinematic distance of 4.9 kpc reported in previous studies.}
The proper motion of IRAS 23385+6053 is obtained to be ($\mu_{\alpha}\cos{\delta}$,$\mu_{\delta}$)=($-3.73\pm0.53$, $-2.0{7}\pm0.73$) mas yr$^{-1}$. 
{Based on VLBI astrometry result, we derived the physical properties of molecular clouds in which H$_2$O masers have been detected, including IRAS 23385+6053 in the Cepheus and Cassiopeia region.}
{We discuss the line-of-sight structures of the giant molecular clouds using the trigonometric distances obtained from the H$_{2}$O maser sources.}
{Our results suggest that molecular clouds in the Perseus arm extend over {approximately $2$ kpc} at the Cepheus and Cassiopeia region.}
\end{abstract}

\section{Introduction} \label{sec:intro}

\begin{figure}[h]
\begin{center}
 \includegraphics[width=8.5cm]{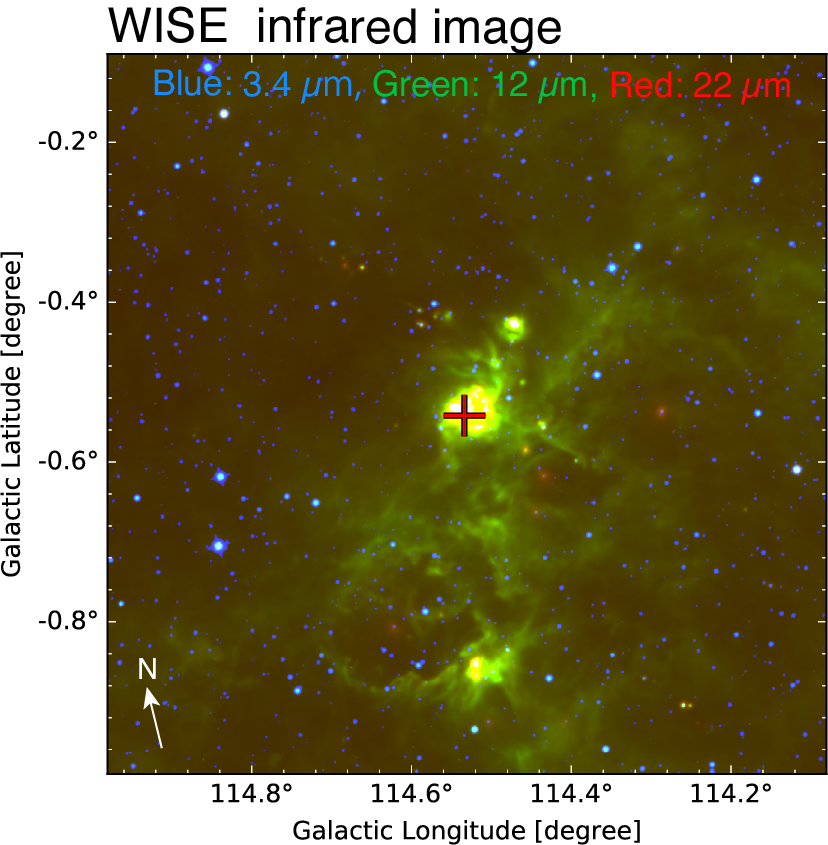}
 \end{center}
\caption{{{A} three-color composite infrared image of IRAS 23385+6053 obtained by the WISE satellite. Blue, green, and red indicate $3.4\ \mu$m, $12\ \mu$m, and $22\ \mu$m, respectively. The red cross mask shows the position of the H$_2$O maser source associated with IRAS 23385+6053. Alt text: The WISE three-color composite image of IRAS 23385+6053.}
\label{wise}}

\end{figure}

Giant molecular clouds (GMCs) have been studied by CO surveys of the Milky Way using single-dish radio telescopes since the 1970s \citep{2001ApJ...547..792D,2004ASPC..317...59M,2015ARA&A..53..583H}. 
GMCs have ideal environments, with massive and dense gas, for the formation of high-mass stars and cluster formation (e.g., \citealp{2003ARA&A..41...57L,2007ARA&A..45..565M,2018ARA&A..56...41M,2019ARA&A..57..227K}), and their evolution and formation {mechanisms have} been widely studied in the Milky Way and Local Group galaxies \citep{2007prpl.conf...81B,2010ARA&A..48..547F,2014prpl.conf....3D,2023ASPC..534....1C}.
In the Milky Way, we can obtain the kinematic distance to molecular clouds using the {local standard of rest (LSR)} velocity of the HI, CO, and/or recombination line data, assuming the Galactic rotation {curve}. 
{However, since the peculiar motion (i.e., non-circular motion) of the Galactic disk can lead to uncertainties of $\pm$(2--3) kpc in the kinematic distance, it is difficult to clarify the three-dimensional structure of GMCs \citep{2009ApJ...706..471B}.}
{Trigonometric parallax measurements are important for directly revealing the precise distances between multiple molecular clouds and their overall three-dimensional distribution.}
{The very long baseline interferometry (VLBI) astrometry observations allow us to measure the} annual parallaxes and proper motions of maser sources associated with young massive stars embedded in molecular clouds (e.g.,\citealp{2007PASJ...59..889H,2007PASJ...59..897H,2011PASJ...63...45A,2012PASJ...64..108S,2015PASJ...67...65N,2015PASJ...67...68N,2015ApJ...800....2H,2016MNRAS.460..283B,2016MNRAS.460.1839C,2019PASJ...71..113K,2020PASJ...72...55O}).

{The VLBI Exploration of Radio Astrometry (VERA) is {a Japanese VLBI network consisting of four radio telescopes located in} Mizusawa, Iriki, Ishigaki Island, and Ogasawara Island.
{The two scientific goals of the VERA project are to reveal the three-dimensional structure of the Milky Way from the annual parallaxes of maser sources, and to reveal the density distribution of dark matter from the Milky Way's rotation curve \citep{2003ASPC..306..367K,Omodaka2009,2012PASJ...64..136H,2020PASJ...72...50V,2025galaxies13.120}. }

{The Cepheus and Cassiopeia region {consists of massive star-forming regions located in the outer Galaxy at Galactic longitudes between $l=\timeform{107D}$ and $l=\timeform{116D}$ \citep{1997ApJS..110...21Y,1998ApJS..115..241H,2000ApJ...537..221U,2022MNRAS.509...68W,2025MNRAS.538..198R}.}
 {The Local arm, the Perseus arm, and the Outer spiral arm are present in the line of sight of this region \citep{2014PASJ...66..104C,2015PASJ...67...69S}.}
IRAS 23385+6053 (= Mol 160) is a young massive star-forming region located {in} the Cepheus and Cassiopeia region \citep{1998ApJ...505L..39M,2003A&A...407..237T,2004A&A...414..299F,2008A&A...487.1119M,2012ApJ...745..116W,2019A&A...627A..68C,2025ApJ...978L..46N}.
Figure \ref{wise} {is a} three-color composite image of IRAS 23385+6053 obtained {using archived data from} the Wide-field Infrared Survey Explorer (WISE) \citep{2010AJ....140.1868W}. Blue, green, and red present the $3.4\ \mu$m, $12\ \mu$m, and $22\ \mu$m, respectively. 
The red cross mark shows the position of the H$_2$O maser source embedded young massive proto-stars.
{Recently, James Webb Space Telescope (JWST) revealed the outflow properties, accretion signatures of embedded massive proto-stars, and detected icy organic molecules as part of the JOYS (JWST Observations of Young protoStars) project \citep{2023A&A...673A.121B,2023A&A...679A.108G,2024A&A...683A.124R,2024A&A...683A.249F}}
The kinematic distance {has been} calculated to be 4.9 kpc \citep{1998ApJ...505L..39M}, {but there are large uncertainties in this value, and} {no maser parallax measurements are reported} for this source (see Section 2 in \citealp{2023A&A...679A.108G}). {Furthermore, it is not clarified the precise locations of molecular clouds inside the Perseus arm at the Cepheus-Cassiopeia region.}
{In this paper, {we used VERA to observe} H$_2$O maser sources {associated with} IRAS 23385+6053 and {newly measured their annual parallax.} We also {discuss the three-dimensional structure of GMCs in the Cepheus and Cassiopeia region} based on VLBI astrometry results obtained by VERA and {the} Very Long Baseline Array (VLBA).}

{This paper is structured as follows: section 2 introduces VLBI observations and archival data sets; section 3 presents the {observational} results of annual parallax and proper motion of H$_2$O maser sources; section 4 presents the physical properties of molecular clouds in the Cepheus and Cassiopeia region, and section 5 {provides a summary} of this paper.}

\section{Observations} 
\label{sec:intro}
\subsection{VLBI observations and data reduction}

\begin{table}[hbtp]
\caption{VERA Observations of IRAS 23385+6053}
 \begin{center}
 \begin{tabular}{llccccccc}
\hline 
\hline 
 & &  \multicolumn{4}{c}{$T_{\rm sys}$ [K]} \\
 \cline{3-6}
ID & Date &MIZ &IRK &OGA &ISG \\
\hline
{A} & 2017 February 21  &103.7 &137.4 &267.5 &240.0 \\
{B} & 2017 April 23  &130.8 &144.4 &239.7 &340.9 \\
{C} & 2017 May 20  &152.0 &151.5 &176.1 &756.5 \\
{D} & 2017 June 5  &111.9 &319.4 &948.5 &517.5 \\
{E} & 2017 September 28   &116.8 &156.6 &547.4 &389.6 \\
{F} & 2017 November 2  &161.8 &176.2 &439.3 &2934.5 \\
{G} & 2017 December 2   &105.7 &190.1 &283.9 &1139.2 \\
{H} & 2018 January 6  &294.0 &135.4 &156.3 &531.2 \\
{I} & 2018 February 15  &94.8 &270.1 &271.9 &241.0 \\
{J} & 2018 March 9  &305.4 &126.6 &264.3 &176.2 \\
{K} & 2018 April 2  &123.1 &195.8 &175.5 &381.4 \\
{L} & 2018 June 5  &196.9 &548.6 &504.4 &572.7 \\
\hline
\end{tabular}\label{table:observation}
\end{center}
{Note. VERA comprises four radio-telescopes in Mizusawa (MIZ), Iriki (IRK), Ogasawara (OGA), and Ishigaki island (ISG) stations. The detection of 22 GHz H$_{2}$O maser was confirmed in all observations. The average system noise temperature for each site is listed.}
\end{table}

We conducted VLBI observations of the H$_{2}$O masers associated with IRAS 23385+6053 at approximately 1-2 month intervals from February 2017 to June 2018. 
The rest frequency of the H$_{2}$O maser with the 6$_{13}$-5$_{23}$ transition is 22.235080 GHz. Details of the observations are summarized in Table \ref{table:observation}. 
VERA is a VLBI astrometry project that reveals the three-dimensional structure of the Milky Way. VERA consists of four 20-m radio telescopes located in Mizusawa, Iriki, Ogasawara, and Ishigaki in Japan \citep{2003ASPC..306..367K,Omodaka2009,2020PASJ...72...50V,2025galaxies13.120}. 
To detect proper motion and annual parallax through calibration of atmospheric phase fluctuation, each telescope is equipped with a dual-beam receiving system  \citep{2000SPIE.4015..544K,2008PASJ...60..935H}, which allows simultaneous observation of the target maser source and the reference radio source. 
As a reference source, we observed J2339+6010, whose coordinates are $(\alpha_{\rm J2000},\delta_{\rm J2000})=(\timeform{23h39m21.s1252}, +\timeform{60D10'11."850})$ \citep{2025ApJS..276...38P}.
During the observations, J2339+6010 was detected with an average flux density of 0.172$\pm$0.028 Jy~beam$^{-1}$.
The separation and position angles (P.A.) of the phase reference source J2339+6010 relative to the maser source IRAS23385+6053 are 1.0$^{\circ}$ and $-$169.1$^{\circ}$, respectively. The P.A. increases eastward, with 0 degrees in the north.

The radio frequency (RF) signal was amplified by each receiver and downconverted to 4.7 - 7.0 GHz via a fixed 1st LO at 16.8 GHz. The intermediate frequency (IF) signal was then mixed down again to a baseband frequency between 0 and 512 MHz. The baseband signal was converted to a digital signal via a 2-bit analog-to-digital converter with a sampling rate of 1024 MHz. The digital signal was filtered with the VERA digital filter \citep{2005PASJ...57..259I} and then recorded at 1024 Mbps, resulting in a total bandwidth of 256 MHz. The data was divided into 16 IFs with each bandwidth of 16 MHz. One IF was assigned to a spectral line source (i.e., maser source),
and the other 15 IFs were assigned to continuum sources (e.g., the phase reference source J2339+6010).
These data were correlated using the Mizusawa software correlator installed at the Mizusawa Campus of the National Astronomical Observatory of Japan \footnote{Mizusawa software correlator: \url{https://www.miz.nao.ac.jp/veraserver/system/fxcorr-e.html}}, and the accumulation period for the correlation process was set to 1 second.
In the correlation processing of IRAS23385+6053, the 16~MHz data was divided into 512~channels, resulting in a velocity resolution of 0.42 km~s$^{-1}$. For continuum data, each IF was divided into 30~channels.


Data reduction was performed with the Astronomical
Image Processing System developed by NRAO (AIPS; \citealp{1996ASPC..101...37V}). Standard data analysis methods were applied to amplitude and phase calibration of the raw data (i.e., visibility), referencing previous VERA publications (e.g., see \citealp{2020PASJ...72...52N}).
The instrumental delay calibration of the VERA dual beam system was performed by the horn-on-dish method \citep{2008PASJ...60..935H}. The tropospheric delay was calibrated using the tropospheric zenith delay measured by the Global Positioning System (GPS) onboard each VERA site.
Ionospheric delays were calibrated using the CODE (the Center for Orbit Determination in Europe) Global Ionosphere Map (GIM)\footnote{Downloaded from \url{ftp://ftp.aiub.unibe.ch/CODE/YYYY/COD0OPSFIN_YYYYDDD0000_01D_01H_GIM.INX.gz} where YYYY and DDD should be replaced by the year and the day of year, respectively.} \citep{2009JGeod..83..263H}.
Earth Orientation Parameters (EOP) referenced C04 2014 solution\footnote{\url{ftp://hpiers.obspm.fr/eop-pc/eop/eopc04_14/eopc04.62-now}}.
Maser images acquired by the AIPS task \texttt{IMAGR} were fitted by a two-dimensional Gaussian function, and the peak position of each maser spot was recorded. A maser feature consists of similarly positioned maser spots with continuously varying velocities.
The recorded positions were used to derive the parallax and proper motion of individual maser features.

\subsection{The archival data}
{We utilized the $^{12}$CO $J$~=~1--0 line data obtained by the Five College Radio Astronomy Observatory (FCRAO) outer Galactic plane survey using the 14-m radio telescope\citep{1998ApJS..115..241H,2000AAS...197.0508B}.
The cube data are taken from the Canadian Galactic Plane Survey website\footnote{https://www.cadc-ccda.hia-iha.nrc-cnrc.gc.ca/en/cgps/}{}\citep{2003AJ....125.3145T}.
The spatial resolution of the cube data is convolved to be $\timeform{100.44"}$.
The final cube data has the voxel size of $(l,b,v)=(\timeform{18"},\timeform{18"}, {\rm 0.82\ km s^{-1}})$. 
The root mean square noise level is $\sim 0.14$ K at the brightness temperature scale.
We also used the archival infrared images data obtained by the WISE satellite \citep{2010AJ....140.1868W}. The WISE data are taken from the SkyView web page \citep{1998IAUS..179..465M}\footnote{https://skyview.gsfc.nasa.gov/current/cgi/query.pl}}

\section{Results}
\subsection{The annual parallax of IRAS 23385+6053}

\begin{table*}[hbtp]
\setlength{\tabcolsep}{3pt}
  \caption{Results of the parallax and the proper motion of the H$_{2}$O maser features associated with IRAS 23385+6053}
  \label{table:data_type}
  \begin{center}
  \small
  \begin{tabular}{cccccccccccccc}
\hline 
\hline 
Feature&{Detection epoch ID}&\multicolumn{4}{c}{Offset}&&\multicolumn{4}{c}{Proper motion}&Parallax&LSR velocity range\\
&&\multicolumn{4}{c}{(mas)}&&\multicolumn{4}{c}{(mas year$^{-1}$)}&(mas)&(km s$^{-1}$)&\\
\cline{3-6}\cline{8-11}&&$X$&$\sigma_{X}$&$Y$&$\sigma_{Y}$&&$\mu_{x}$&$\sigma_{\mu_{x}}$&$\mu_{y}$&$\sigma_{\mu_{y}}$&&\\ 
\hline

1&{A\underline{B}CD\texttimes\texttimes\texttimes\texttimes\texttimes\texttimes\texttimes\texttimes}&$-$25.{8}2&0.0{1}& 2{4.91}&0.0{1}&&$-$4.37&0.10&$-$0.61&0.13&$-$&$-$52.0 $\sim$ $-$50.8\\
2&{A\underline{B}C\texttimes\texttimes\texttimes\texttimes\texttimes\texttimes\texttimes\texttimes\texttimes}&$-$27.21&0.05& 21.28&0.05&&$-$4.51&0.35&$-$1.77&0.26&$-$&$-$49.9 $\sim$ $-$49.5\\
3&{\texttimes\texttimes\texttimes\texttimes\texttimes\texttimes\texttimes\texttimes I\underline{J}K\texttimes}&$-$31.20&0.02& 31.50&0.02&&$-$4.73&0.59&$-$1.25&0.22&$-$&$-$59.2 $\sim$ $-$58.8\\

4&{ABCD\underline{E}FGHI\texttimes\texttimes\texttimes}&$-$6.{50}&0.0{1}& 23.{38}&0.0{1}&&$-$3.38&0.17&$-$4.00&0.54&0.470$\pm$0.082&$-$55.4 $\sim$ $-$53.7\\

5&{ABC\texttimes\underline{E}FGHI\texttimes\texttimes\texttimes}&18.{49}&0.0{1}& $-$15.{72}&0.0{1}&&$-$3.06&0.20&$-$1.97&0.24&0.446$\pm$0.094&$-$49.1 $\sim$ $-$44.0\\

6&{\texttimes\texttimes\texttimes\texttimes EF\underline{G}HIJ\texttimes\texttimes}&18.{97}&0.02&  $-$1{0.85}&0.02&&$-$4.22&0.63& $-$4.93&1.03&$-$&$-$40.7 $\sim$ $-$33.9\\

7&{A\underline{B}C\texttimes\texttimes\texttimes\texttimes\texttimes\texttimes\texttimes\texttimes\texttimes}&29.7{0}&0.0{2}& 20.{39}&0.0{2}&&$-$3.44&0.60&0.06&0.41&$-$&$-$47.4 $\sim$ $-$46.6\\

8&{\texttimes\texttimes\texttimes\texttimes\texttimes\texttimes\texttimes HI\underline{J}KL}&27.46&0.01& 25.9{5}&0.01&&$-$3.21&0.10&$-$0.50&0.28&$-$&$-$58.0 $\sim$ $-$56.7\\

9&{\texttimes\texttimes\texttimes\texttimes\texttimes\texttimes\texttimes\texttimes IJ\underline{K}L}&25.{50}&0.04& $-$27.06&0.03&&$-$3.21&0.14&$-$4.25&0.19&$-$&$-$49.1 $\sim$ $-$46.1\\

10&{A\underline{B}CD\texttimes\texttimes\texttimes\texttimes\texttimes\texttimes\texttimes\texttimes}&$-$10.75&0.01& 16.13&0.02&&$-$3.21&0.14&$-$1.44&0.37&$-$&$-$60.1 $\sim$ $-$59.2\\
\hline 
\multicolumn{2}{l}{Combined fit (features 4 and 5)}&$-$&$-$&$-$&$-$&&$-$&$-$&$-$&$-$&0.460$\pm$0.061&$-$\\
\hline
\label{table:spot}
  \end{tabular}
  \end{center}
{Note. {See Table \ref{table:observation} for dates of detection epoch IDs. A cross means it was not detected.} $X$ and $Y$ are the position offsets relative to the phase reference source in right ascension ($\alpha$cos$\delta$) and declination ($\delta$), respectively. {The epoch to which the position offset refers is underlined.} $\sigma_X$ and $\sigma_Y$ are the formal errors of the position offsets. The position origin is ($\alpha$cos$\delta$, $\delta)_{\rm J2000.0}$ = $(\timeform{23h40m54s.5224}, +\timeform{61D10'28''.080}$). $\mu_x$ and $\mu_y$ are proper motion components in right ascension and declination, respectively. $\sigma_{\mu_x}$ and $\sigma_{\mu_y}$ are the errors of the proper motion components. The parallax fit was performed only on the two most long-lived (359 days) maser features.}
\end{table*}

\begin{figure}[h]
 \includegraphics[width=8.6cm]{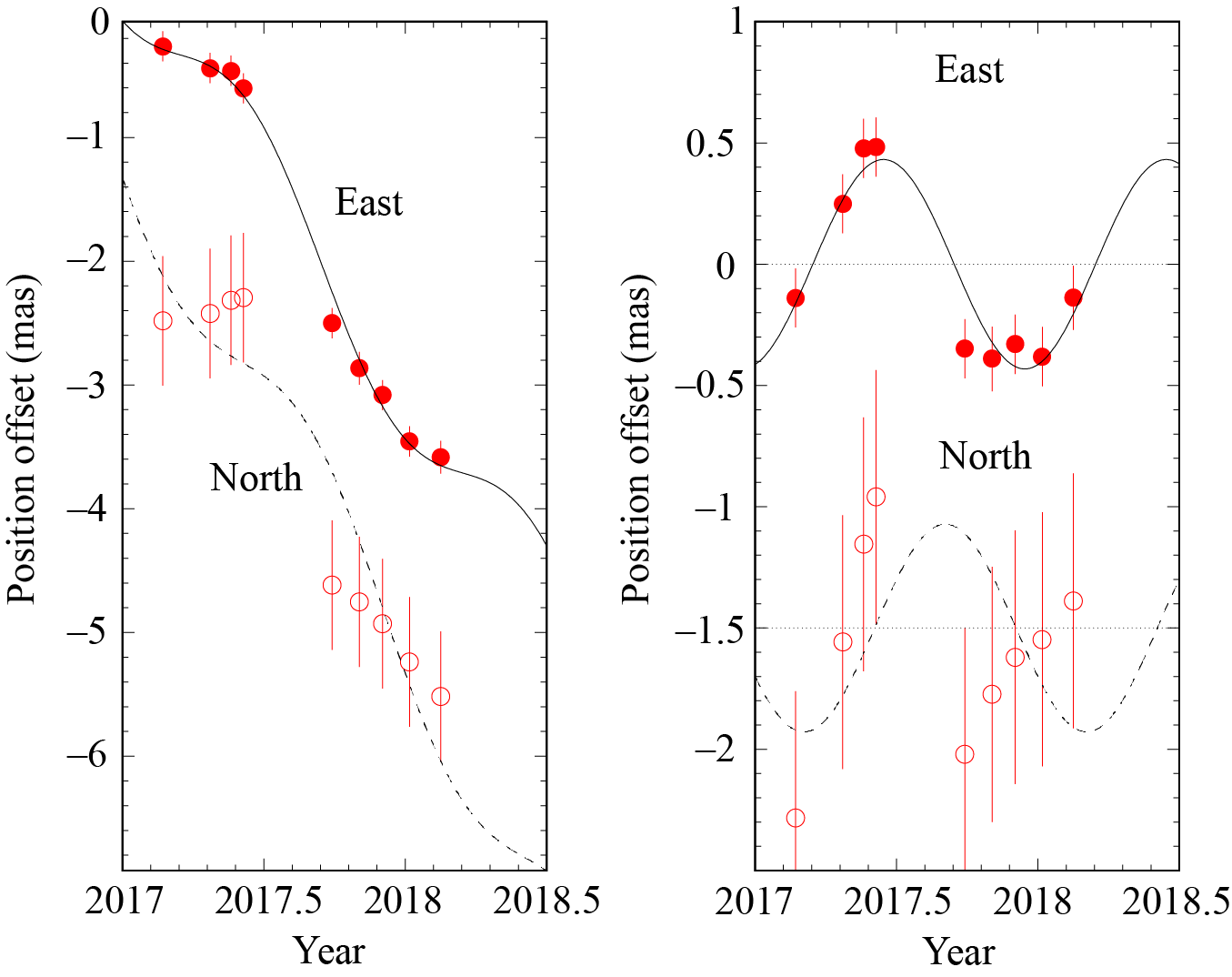}
\caption{Parallax and proper motion results for the maser feature 4 (see Table \ref{table:data_type}). \textbf{(Left)} Position offsets in the east (filled circles) and north (open circles) directions relative to the phase reference J2339+6010 are shown as a function of time. For clarity, the northerly data is plotted offset from the easterly data. The solid and dashed curves represent the results of model fitting to the easterly and northerly data, respectively. \textbf{(Right)} Same as left, but with the proper motion components subtracted. Only the parallax result is shown.}
{Alt text: Two curve graphs. In both panels, the x-axis shows the years from 2017 to 2018.5. The model that best fits the observed data is shown in each panel.}
\label{feature2}
\end{figure}

 {10 maser features were detected in IRAS 23385+6053, with velocities ranging from $V_{\rm LSR}=-60.1$ km s$^{-1}$ to $-33.9$ km s$^{-1}$.
Table \ref{table:spot} presents the proper motion and parallax results for the maser features. {The differences in the number and timing of detection of each maser feature are thought to be due to (1) the variability and lifetime of each maser feature, and (2) the noise level of each observational data.} Note that the parallaxes were determined using the two longest-lived maser features (359 days), and the proper motions were determined using maser features that were confirmed to be detected for three epochs or more. In VLBI astrometry, systematic errors are larger than statistical errors \citep{2014ARA&A..52..339R}, so when determining individual proper motions and annual parallaxes, a constant systematic error was introduced so that the reduced chi square value of the least-squares fitting approaches one.
Figure \ref{feature2} shows the annual parallax and proper motion results for the maser feature 4. Based on a combined fit of maser features 4 and 5, the annual parallax of IRAS 23385+6053 is $0.460 \pm 0.086$ mas, which corresponds to a distance of $2.17^{+0.50}_{-0.34}$ kpc. The error of the parallax is multiplied by $\sqrt{N}$, where $N$ is the number of maser features. This is because maser features are expected to be affected by the same systematic errors (e.g., atmospheric delay residuals).}

\subsection{The systemic proper motion of IRAS 23385+6053}

Observed each proper motion consists of a systemic (e.g., galactic rotation) and an internal motion of each maser feature (e.g., outflow). If we assume that the internal motion is random, we can obtain the systemic proper motion of IRAS 23385+6053 by averaging the proper motion results. The averaged proper motion components are $(\mu_{\alpha}\cos{\delta},\mu_{\delta})=$ ($-3.73\pm0.53$, $-2.0{7}\pm0.73$) mas yr$^{-1}$. We adopt these values for the central star that excites the maser. To account for this uncertainty, we added $\pm$5 km s$^{-1}$ in quadrature to the standard error of each motion component. The uncertainty of 5 km s$^{-1}$ is consistent with the velocity dispersion of a typical molecular cloud \citep{1985ApJ...295..422C}. Figure \ref{motion} shows the spatial distribution of H$_2$O maser features. The maser features are distributed over $\sim$80 $\times$ $\sim$80 mas, which corresponds to $\sim$174 $\times$ $\sim$174 AU at a distance of 2.17 kpc. 
The arrows represent the internal motion of the maser features.


\begin{figure}[htb]
\begin{center}
\includegraphics[width=11cm]{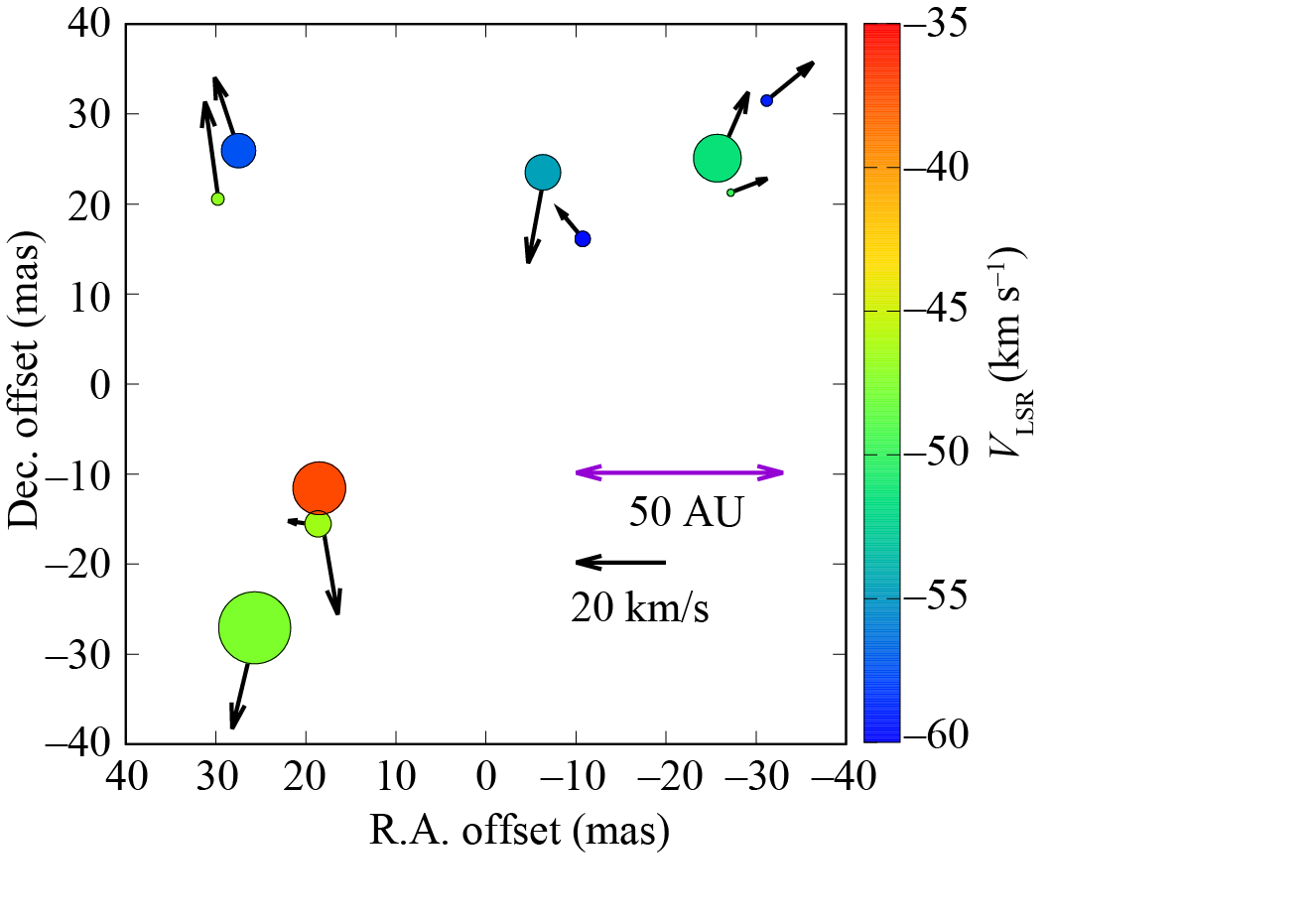}
 \end{center}
\caption{Distribution and internal motion of H$_{2}$O maser features in IRAS 23385+6053. The map origin is $(\alpha_{\rm J2000},\delta_{\rm J2000}) = (\timeform{23h40m54.s5224}, \timeform{61D10'28"080})$. The color of each circle shows LSR velocity. The arrows represent the internal motion vectors. The scales of 50 AU and 20 km s$^{-1}$ at a distance of 2.17 kpc are shown in the bottom right of the map, respectively. The size of each circle is proportional to the peak flux density, with the minimum and maximum flux densities being 0.8 Jy and 8.0 Jy, respectively.}{Alt text: Distribution of colored circles with motion vectors. The x-axis shows R.A. offset from 40 to $-$40 mas, and the y-axis represents Dec. offset from $-$40 to 40 mas. The LSR velocity range (color bar range) is from $-$60 to $-$35 km s$^{-1}$. The scales of 50 AU and 20 km s$^{-1}$ correspond to 23 mas and 10 mas on the map, respectively.}
\label{motion}
\end{figure}

\subsection{The peculiar motion and the rotation velocity in the Galactic disk}
The peculiar motion is a deviation from the circular motion of the Galactic disk (i.e., non-circular motion). The peculiar motion of IRAS 23385+6053 was estimated based on \citet{2015PASJ...67...69S, 2020PASJ...72...53S}. 
The LSR velocity of IRAS 23385+6053 is $V_{\rm~LSR} = -49.1\pm 4.7$ km s$^{-1}$ based on $^{12}$CO ($J$ \~=~1--0) observations \citep{1989A&AS...80..149W}. The Galactic constants $R_{0}$ = 7.92~kpc and $\Theta_{0}$ = 239~km~s$^{-1}$ \citep{2020PASJ...72...50V} were assumed, where $R_{0}$ is the distance to the Galactic center and $\Theta_{0}$ is the circular velocity at the solar position. The solar motion value from \cite{2010MNRAS.403.1829S}, ($U_{\odot}$, $V_{\odot}$, $W_{\odot}) = (11.1, 12.24, 7.25)$ km s$^{-1}$, was adopted. Our parallax and proper motion results for IRAS 23385+6053 were used to determine the peculiar motion. As a result, the peculiar motion of the source was determined to be $(U_{\rm{s}},V_{\rm{s}},W_{\rm{s}}$) = (25$\pm$8, $-$12$\pm$7, $-$2$\pm$8) km s$^{-1}$. Here, $U_{\rm{s}}$, $V_{\rm{s}}$, and $W_{\rm{s}}$ point towards the Galactic center, Galactic rotation, and north Galactic pole, respectively.
The rotation velocity of IRAS 23385+6053 was determined to be $\Theta_{s}$ = 227$\pm$7 km~s$^{-1}$, which is smaller than $\Theta_{0}$ = 239 km~s$^{-1}$ \citep{2020PASJ...72...50V}. 

All results of the IRAS 23385+6053 observations are summarized in Table \ref{table:obs_results}.

\begin{table*}[hbtp]
  \caption{Observation results of IRAS 23385+6053}
  \label{table:obs_results}
  \begin{center} 
  \begin{tabular}{ll}
    \hline \hline
    Annual parallax\footnotemark[$*$]& $0.460 \pm 0.086$~mas\\
    Distance & $2.17^{+0.50}_{-0.34}$~kpc\\
    Systemic proper motion\footnotemark[$\dag$]& $(\mu_{\alpha}\cos{\delta},\mu_{\delta})=(-3.73 \pm 0.53,-2.07 \pm 0.73)$~mas~yr$^{-1}$\\
    Peculiar (non-circular) motion& ($U_{s}$,$V_{s}$,$W_{s}$) = $(25 \pm 8, -12 \pm 7, -2 \pm 8)$~km~s$^{-1}$\\
    Systemic LSR velocity\footnotemark[$\ddag$] & $-49.1 \pm 4.7$ km s$^{-1}$ \\
    Rotation velocity & $\Theta_{s}=227\pm7$~km~s$^{-1}$\\
    \hline\label{table:I23385}
\end{tabular}
\end{center}
\begin{flushleft}
\footnotesize
$^{*}$The error of the combined fit (see Table \ref{table:data_type}) is increased by a factor of two since the two maser features may be subject to a common systematic error (i.e., atmospheric delay residuals).\\
$^{\dag}$ Each error is the root sum square of the statistical and systematic (0.49 mas yr$^{-1}$) errors (see the text).\\
$^{\ddag}$ \cite{1989A&AS...80..149W}
\end{flushleft}
\end{table*}

\section{Discussion}
\subsection{The Cepheus-Cassiopeia GMCs}
\begin{figure*}[h]
\begin{center}
\includegraphics[width=14cm]{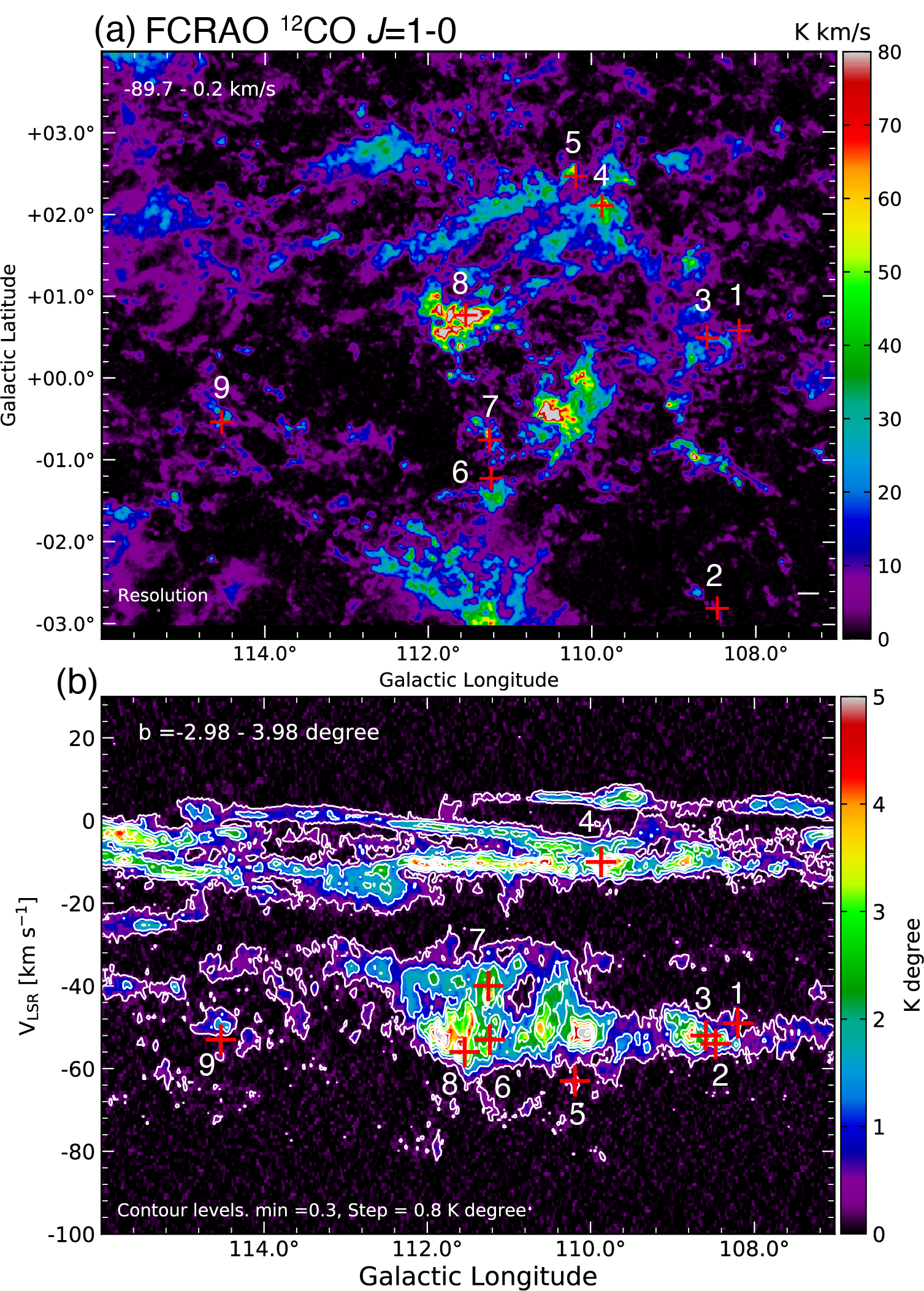}
\end{center}
\caption{{(a) The $^{12}$CO $J$~=~1--0 integrated intensity map obtained by the outer Galactic plane CO survey \citep{1998ApJS..115..241H}. The integrated velocity range is from $-90$ km~s$^{-1}$ to $0$ km~s$^{-1}$. (b) The $^{12}$CO $J$~=~1--0 longitude-velocity diagram {for} the integrated latitude range from $\timeform{-3.0D}$ to $\timeform{+4.0D}$. The cross marks show the H$_2$O maser sources whose {distances have} been measured by VERA and VLBA \citep{2019ApJ...885..131R,2020PASJ...72...50V}. The lowest contour level and interval are 0.3 K degree and 0.8 K degree, respectively. {The source names corresponding to the numbers (IDs) are shown} in Figure \ref{fig:h20} and Table \ref{table:phys}. Alt text: The 12-CO integrated intensity map and longitude-velocity diagram.}
\label{lv}}
\end{figure*}

\begin{figure*}[h]
\includegraphics[width=16cm]{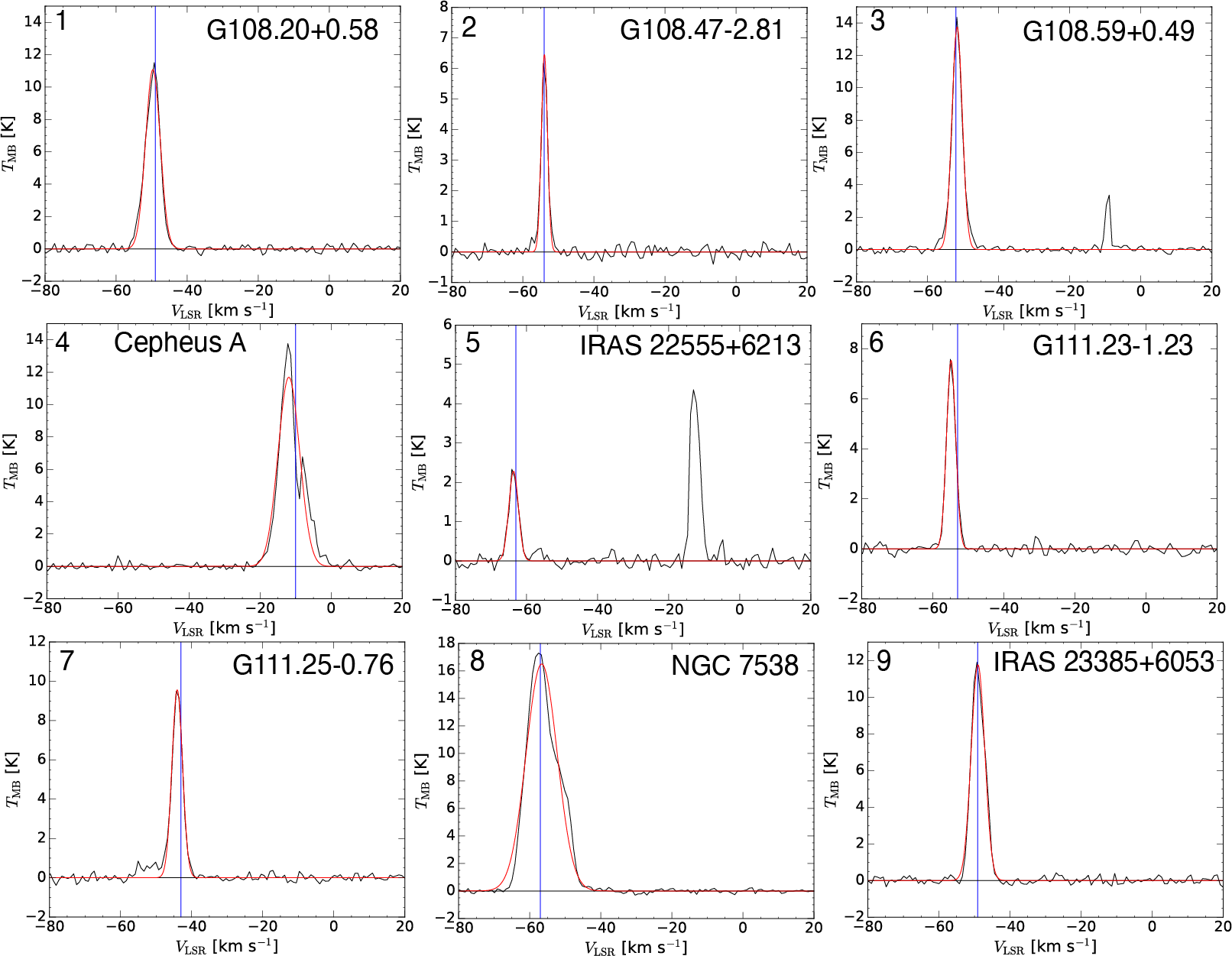}
\caption{{CO spectra containing each H$_2$O maser {source.} Red lines show the fitted results of the Gaussian function fitting}. The vertical blue line indicates the peak velocity of each H$_2$O maser source. Alt text: CO spectrum associated with each H$_2$O maser source.}
\label{fig:h20}
\end{figure*}

\begin{figure*}[h]
\begin{center}
\includegraphics[width=14cm]{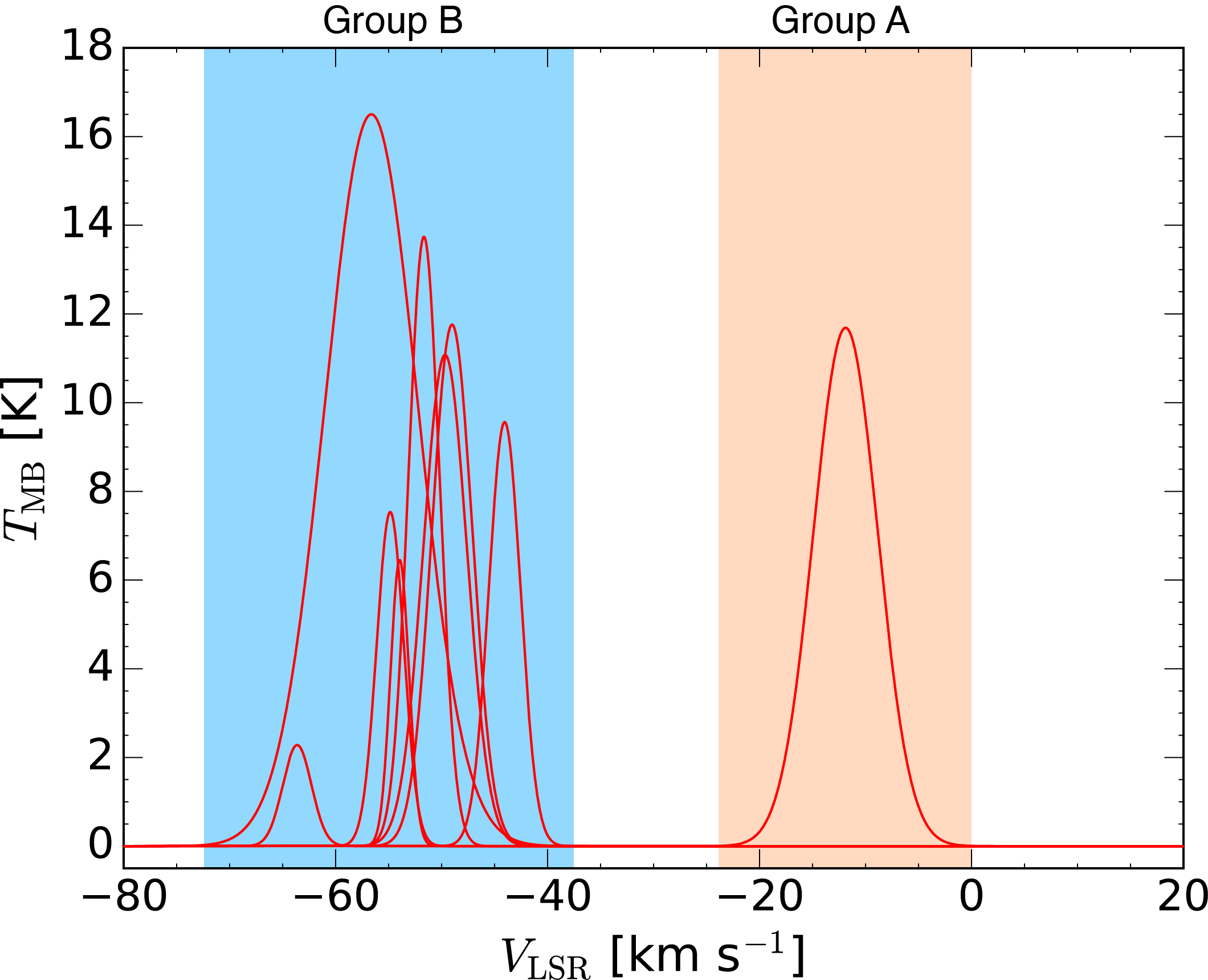}
\caption{The fitting results of the CO spectra {containing H$_2$O maser sources}. Alt text: The fitting results of the CO spectra}
\label{fig:Gauss}
\end{center}
\end{figure*}

{Previous studies of CO observations revealed the large-scale distributions of molecular clouds in the Cepheus-Cassiopeia region including IRAS 23385+6053 \citep{1997ApJS..110...21Y,1998ApJS..115..241H,2000ApJ...537..221U}.
Figures~\ref{lv}(a) and \ref{lv}(b) show the $^{12}$CO $J=$1-0 integrated intensity map and the longitude-velocity diagram obtained by the FCRAO outer Galactic plane survey \citep{1998ApJS..115..241H}.}
We find the two main velocity components with $V_{\rm LSR}=-50$ km s$^{-1}$ and $V_{\rm LSR}=-10$ km s$^{-1}$ in the longitude-velocity diagram. The $-10$ \kms component has uniform velocity distributions, while the $-50$ \kms component concentrates at the Galactic longitude between $l=\timeform{110D}$ and $l=\timeform{113D}$ on the longitude-velocity diagram.
The H$_2$O maser sources are plotted as the cross marks. 
IRAS 23385+6053 is located at $(l,b,v) \sim (\timeform{114.53D},\timeform{-0.54D}, -50\ {\rm km\ s}^{-1})$. The trigonometric distance of other sources were already obtained by VERA and VLBA \citep{2014ApJ...790...99C,2016SciA....2E0878X,2009ApJ...693..406M,2014PASJ...66..104C}.
Figure \ref{fig:h20} shows the CO spectra at the positions of each H$_2$O maser source. All sources of peak {LSR} velocity coincide between the CO spectrum and H$_2$O maser source. 
Therefore, we can consider the molecular clouds {containing H$_2$O maser source are thought to be at same distances as those obtained from the annual parallax measurements using VLBI observations.}

Figure \ref{fig:Gauss} shows the fitting spectra of CO clouds {containing} H$_2$O maser sources.
In the Cepheus and Cassiopeia {region}, CO gas has mainly two velocity components{, mainly} $-10$ \kms and $-70$ \kms to $-40$ \kms, {which we will refer to as "Group A" and "Group B" below.}

\begin{figure*}[h]
\begin{center}
\includegraphics[width=14cm]{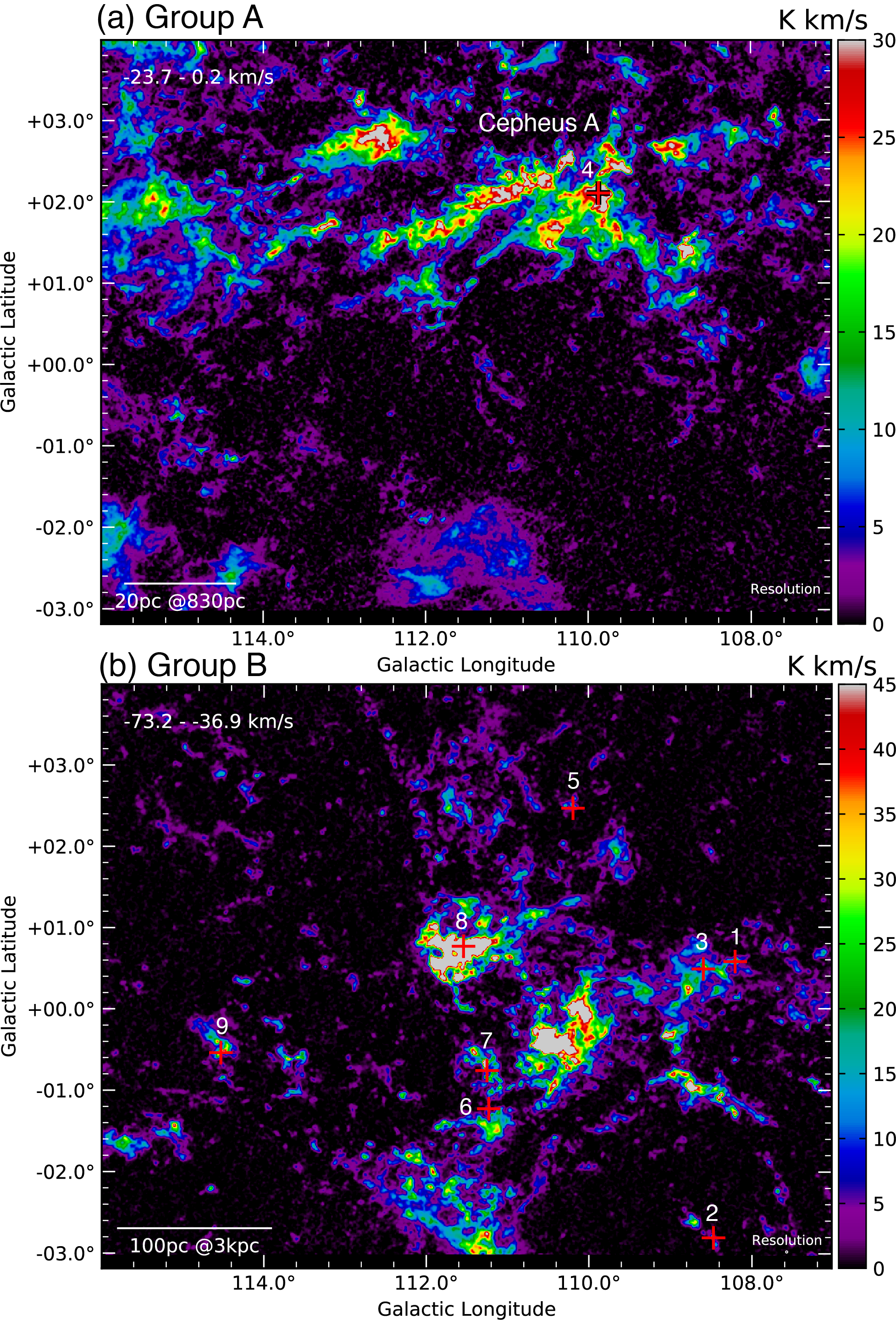}
\end{center}
\caption{(a) The $^{12}$CO $J=$ 1-0 integrated intensity map {for} the velocity range from $V_{\rm LSR}=-24$ km~s$^{-1}$ to $V_{\rm LSR}=0$ km~s$^{-1}$. The red cross mark indicates the position of {the} H$_2$O maser source associated with Cepheus A (No.4). (b) The $^{12}$CO $J=$ 1-0 integrated intensity map {for} the velocity range from $V_{\rm LSR}=-73$ km s$^{-1}$ to $V_{\rm LSR}=-37$ km s$^{-1}$. The red cross marks indicate {the positions} of H$_2$O maser {sources} associated with G108.20+00.58 (No.1), G108.47-02.81 (No.2), G108.59+00.49 (No.3), IRAS 22555+6213 (No.5), G111.23-01.23 (No.6), G111.25-0.77 (No.7), NGC 7538 (No.8), and IRAS 23385+6053 (No.9). Alt text: The 12-CO integrated intensity maps {for} the integrated velocity range corresponding to Cepheus A and Perseus arm.}
\label{cepheus}
\end{figure*}

\subsection{Physical properties of molecular clouds}
{We calculated the physical properties of molecular clouds based on the trigonometric distance obtained by the VLBI astrometry results. The H$_2$ column densities and molecular masses {are calculated} from the $^{12}$CO integrated intensity.
The H$_2$ column density using the CO-to-H$_2$ conversion factor ($X_{\rm CO}$) is given by
\begin{equation}
N({\rm H_2})_{\rm 12x} = X_{\rm CO}\ I_{\rm ^{12}CO}\ {\rm [cm^{-2}]},
\label{XCO}
\end{equation}
where $I_{\rm ^{12}CO}$ is the integrated intensity of $^{12}$CO~$J$~=~1--0.
In this paper, we applied $X_{\rm CO} = 2 \times 10^{20}\ {\rm [cm^{-2}\ (K\ km\ s^{-1})^{-1}]}$ as the standard value of the Galactic disk in the Milky Way \citep{2013ARA&A..51..207B,2020MNRAS.497.1851S,2024MNRAS.527.9290K}.
The size parameter of a molecular cloud is given by
\begin{equation}
S = D \tan (\sqrt{\sigma_l\ \sigma_b})\ [{\rm pc}],
\label{size}
\end{equation}
where $D$, $\sigma_l$, and $\sigma_b$ are the distance from the solar system, the intensity-weighted standard deviation {values} along the Galactic longitude and Galactic latitude {axes}, respectively.
 The total molecular mass is given by
\begin{equation}
M = \mu_{\rm H_2} m_{\rm H} D^2 \sum_{i} \Omega\ N_i({\rm H_2})\ [M_{\odot}],
\label{mass}
\end{equation}
where $\mu_{\rm H_2}$, $m_{\rm H}$, $\Omega$, and $N_i(\rm{H_2})$ {are} the mean molecular weight per hydrogen molecule of 2.8, proton mass of $1.67 \times 10^{-24}$ g, solid angle, and the H$_2$ column density in each pixel, respectively. We derived {size parameters and molecular masses using the $^{12}$CO $J$~=~1--0 line data above the $5\sigma$ noise level}.
The physical parameters of each molecular cloud based on the trigonometric distance are summarized in Table \ref{table:phys}.

\begin{table*}[hbtp]
  \caption{Physical properties of molecular clouds}
  {
  \begin{tabular}{ccccccccccc}
    \hline
     ID &Name & $V_{\rm LSR}^{\rm H_2O}$  & $V_{\rm LSR}^{\rm CO}$ &Trigonometric &$T_{\rm peak}$ & $\sigma_v$ & $S$ & $N({\rm H_2})_{\rm max}$ & Molecular mass\\
     & & [km s$^{-1}$] & [km s$^{-1}$]& distance [kpc] & [K] & [km s$^{-1}$] & [pc] & [cm$^{-2}$] & [$M_{\odot}$] \\
    (1) & (2)& (3) & (4) & (5) & (6) & (7) & (8)& (9)& (10) \\
    \hline \hline
     1& G108.20$+$0.58 & $-49 \pm 5$ & $-49$ & $4.41^{+0.86}_{-0.62}$&$12$& $2.4$ & {$9.2^{+1.8}_{-1.3}$} & $1.1\times10^{22}$ & {$\left(3.8^{+1.6}_{-1.0}\right) \times 10^4$} \\
     2& G108.47$-$2.81& $-54 \pm 5$ & $-56$ &{$3.24^{+0.11}_{-0.10}$} &$9.5$& $1.5$ & {$6.7^{+0.2}_{-0.2}$} & $6.7 \times 10^{21}$ &{$\left(8.3^{+0.6}_{-0.5}\right) \times 10^3$} \\
     3& G108.59$+$0.49 &$-52 \pm 5$ & $-52$ & $2.47^{+0.22}_{-0.19}$&$14$&$2.2$ & {$7.0^{+0.6}_{-0.5}$} & $1.1 \times 10^{22}$ &{$\left(3.1^{+0.6}_{-0.5}\right) \times 10^4$} \\
     4& Cepheus A &  $-10 \pm 3$ & $-9.7$ &$0.83^{+0.02}_{-0.02}$&$22$ & $3.1$ & {$19^{+0.5}_{-0.5}$} & $1.8 \times 10^{22}$ &  {$\left(1.1^{+0.05}_{-0.05}\right) \times 10^5$}\\
     5& IRAS 22555$+$6213 & $-63 \pm 6$ & $-63$&$3.18^{+0.90}_{-0.66}$ & $6.3$& $1.4$ & {$4.0^{+1.1}_{-0.8}$} & $4.8 \times 10^{21}$ &{$\left(3.5^{+2.3}_{-1.3}\right) \times 10^3$} \\
     6& G111.23$-$1.23 &$-53 \pm 10$ & $-55$&$3.33^{+1.23}_{-0.71}$ &$7.6$ & $1.9$ & {$7.9^{+2.9}_{-1.7}$} & $4.9 \times 10^{21}$ &{$\left(2.0^{+1.7}_{-0.8}\right) \times 10^3$} \\
     7& G111.25$-$0.77 &$-43 \pm 5$ & $-44$ &$3.34^{+0.27}_{-0.23}$ &$21$& $2.3$ & ${8.7^{+0.7}_{-0.6}}$ & $1.8 \times 10^{22}$ & {$\left(4.1^{+0.7}_{-0.5}\right) \times 10^4$} \\
     8& NGC 7538 &$-57 \pm 5$ & $-57$ & $2.65^{+0.12}_{-0.11}$&$26$& $3.5$ & {$8.3^{+0.4}_{-0.3}$} & $3.7 \times 10^{22}$  &{$\left(2.0^{+0.2}_{-0.2}\right) \times 10^5$} \\
     9& IRAS 23385$+$6053 & $-49 \pm 5$ & $-49$ &$2.17^{+0.50}_{-0.34}$ & $12$ & $1.4$ & {$5.1^{+1.2}_{-0.8}$} & $1.1 \times 10^{22}$ & {$\left(1.1^{+0.6}_{-0.3} \right)\times 10^4$} \\   
    \hline
    \label{table:phys}
  \end{tabular}}\\
  {\raggedright Notes. --- Columns: (1) ID number (2) Region name (3) {LSR} velocity of a H$_2$O maser source obtained by the VLBI observations \citep{2014ApJ...790...99C,2016SciA....2E0878X,2009ApJ...693..406M,2014PASJ...66..104C}. (4) {LSR} velocity of a molecular cloud at the peak position. (5) Trigonometric distance taken from the VLBI observations. (6) Peak brightness temperature of $^{12}$CO $J=$1--0. (7) Intensity-weighted standard deviation of the {LSR} velocity. (8) Size parameter of a molecular cloud (9) Peak H$_2$ column density assuming the CO-to-H$_2$ conversion factor. (10) Total molecular mass. {The uncertainties in the size parameters and molecular masses are estimated from the uncertainties in the trigonometric distances.}\par}
\end{table*}

Figure \ref{cepheus}(a) shows the integrated intensity map of Group A GMC. 
Group A corresponds to the Cepheus A GMC. 
The size and total molecular mass of GMC are derived to be 100 pc $\times$ 50 pc ($l \times b$) and $\sim 1.1\times 10^5$ $M_{\odot}$, adopting the trigonometric distance of {0.83 kpc at Cepheus A} \citep{2016SciA....2E0878X}.
Figure \ref{cepheus}(b) shows the integrated intensity map of Group B GMC. In this velocity range, 8 H$_2$O maser sources {each with a measured annual parallax, are associated with individual molecular gases} as shown by red crosses. The molecular cloud mass is concentrated in NGC 7538 (No.8), while G108.47-2.81 (NO.2), IRAS 22555+6213 (NO.5), and IRAS 23385+6053 (No.9) are located $\sim 100$--$150$ pc away {from} the main molecular gas component. The total molecular mass is estimated to be $\sim 1.7 \times 10^6\ M_{\odot}$ adopting the distance of 3 kpc, which is the average distance of 8 H$_2$O maser sources presented in this paper.

\begin{figure*}[h]
\begin{center}
\includegraphics[width=13cm]{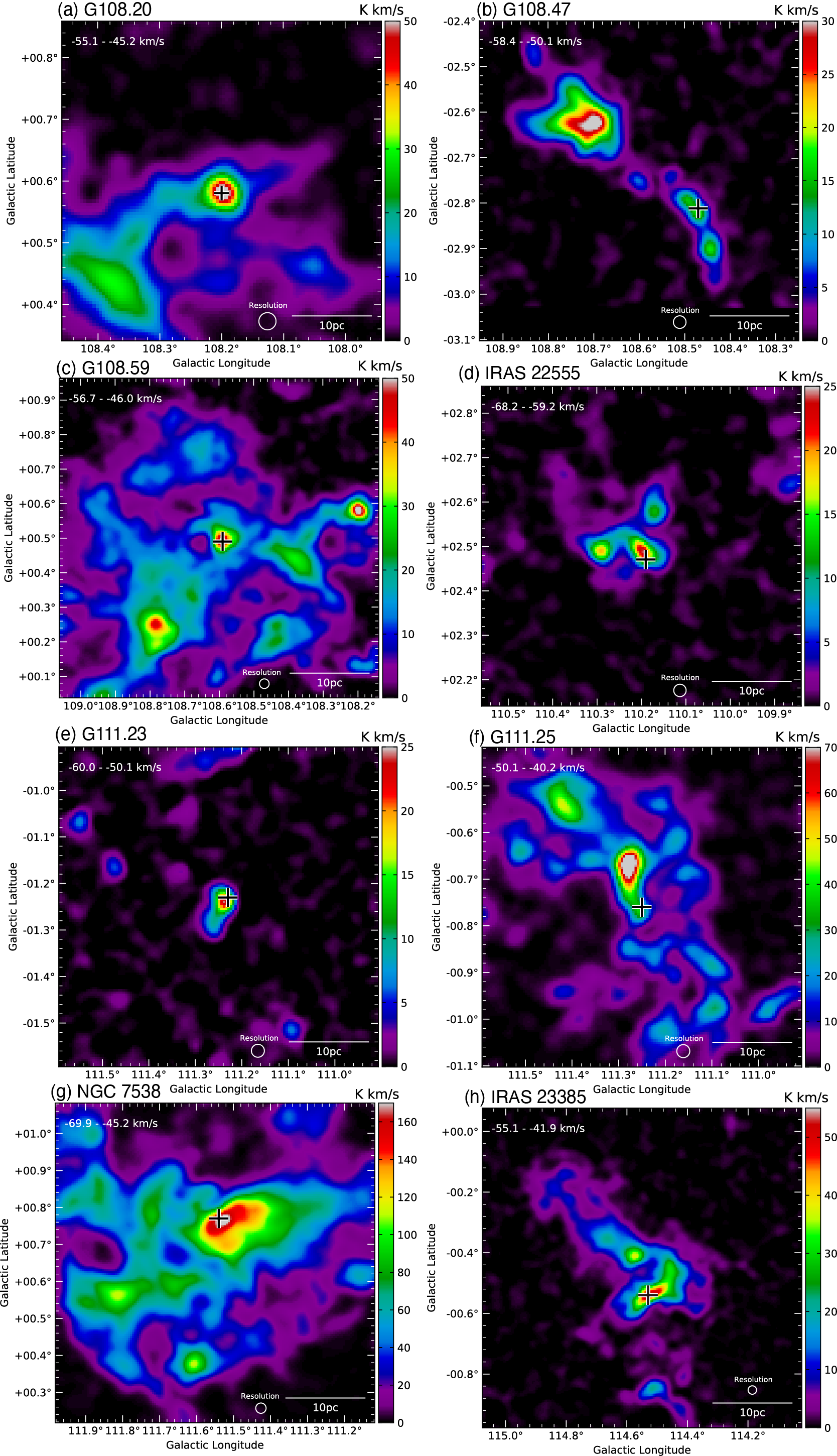}
\end{center}
\caption{The $^{12}$CO $J=$ 1--0 integrated intensity maps of (a) G108.20+0.58, (b) G108.47-2.81, (c) G108.59+0.49, (d) IRAS 22555+6213, (e) G111.23-1.23, (f) G111.25-0.77, (g) NGC 7538, and (h) IRAS 23385+6053. The black crosses show the position of {each} H$_2$O maser source. Alt text: The 12-CO integrated intensity maps focusing on each H$_2$O maser source}
\label{comap}
\end{figure*}

\begin{figure*}[h]
\begin{center}
\includegraphics[width=13cm]{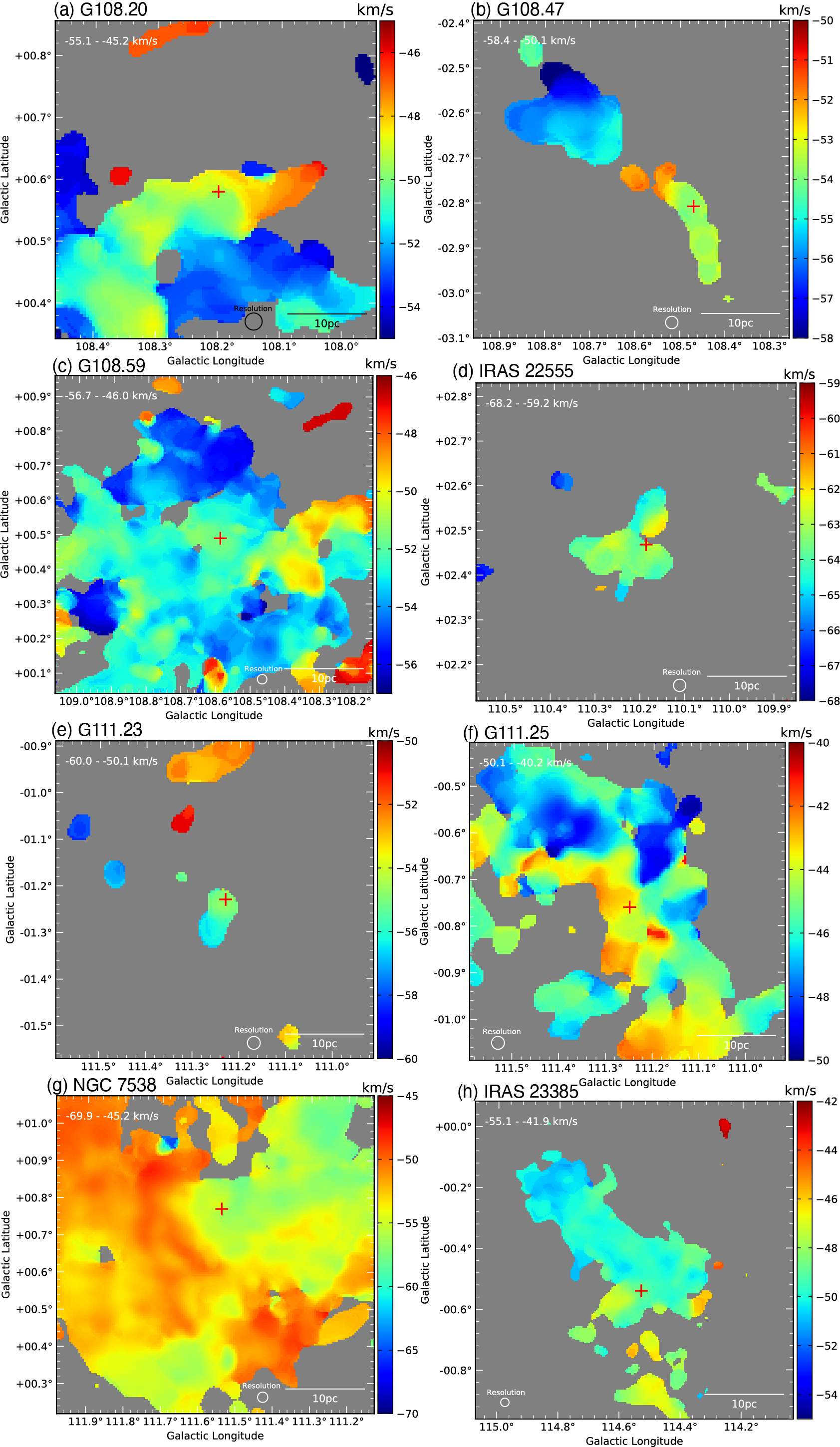}
\end{center}
\caption{The $^{12}$CO $J$~=~1--0 first-moment ($V_c$) maps of (a) G108.20+0.58, (b) G108.47-2.81, (c) G108.59+0.49, (d) IRAS 22555+6213, (e) G111.23-1.23, (f) G111.25-0.77, (g) NGC 7538, and (h) IRAS 23385+6053. The red crosses show the position of {each} H$_2$O maser source. Alt text: The 12-CO first-moment maps focusing on each H$_2$O maser source}
\label{comom1}
\end{figure*}

Figure \ref{comap} presents the integrated intensity maps {focused} on each H$_2$O source {in} Group B.
We can find that {the} molecular gas has a compact distribution {in} G108.20+0.58, IRAS 22555+6213, and G111.23-1.23. On the other hands, G108.47-2.81, G111.25-0.77, and IRAS 23385+6053 has elongated CO distributions.
The filamentary {morphology} of molecular clouds in the Cepheus and Cassiopeia region {may} be explained by Galactic-scale dynamics such as spiral shocks and/or shear motions (e.g., \citealp{2014PASA...31...35D,2020ApJ...896...36T,2022PASJ...74...24K}). 
Figure \ref{comom1} demonstrates the first moment maps focusing on each H$_2$O source in Group B.
The first-moment corresponding to the intensity-weighted mean velocity ($V_c$) {is} calculated as
\begin{eqnarray}
V_{\rm c} &=& {\int T_{\rm B}(v)  v\ dv \over \int T_{\rm B}(v)\ dv},
\label{eq:vc}
\end{eqnarray}
, where $T_{\rm B}$ is the $^{12}$CO brightness temperature and $v$ is the {LSR}} velocity.
Each GMC {with an} H$_2$O maser source {exhibits a} velocity gradient {of about} $\sim$ 5 km s$^{-1}$ to 10 \kms. 
{This result may be caused by internal turbulent motions within the molecular cloud influenced by embedded protostars containing H$_2$O maser sources.}

\subsection{The line-of-sight structure of the Cepheus and Cassiopeia giant molecular cloud}
\begin{figure*}[h]
 \includegraphics[width=15cm]{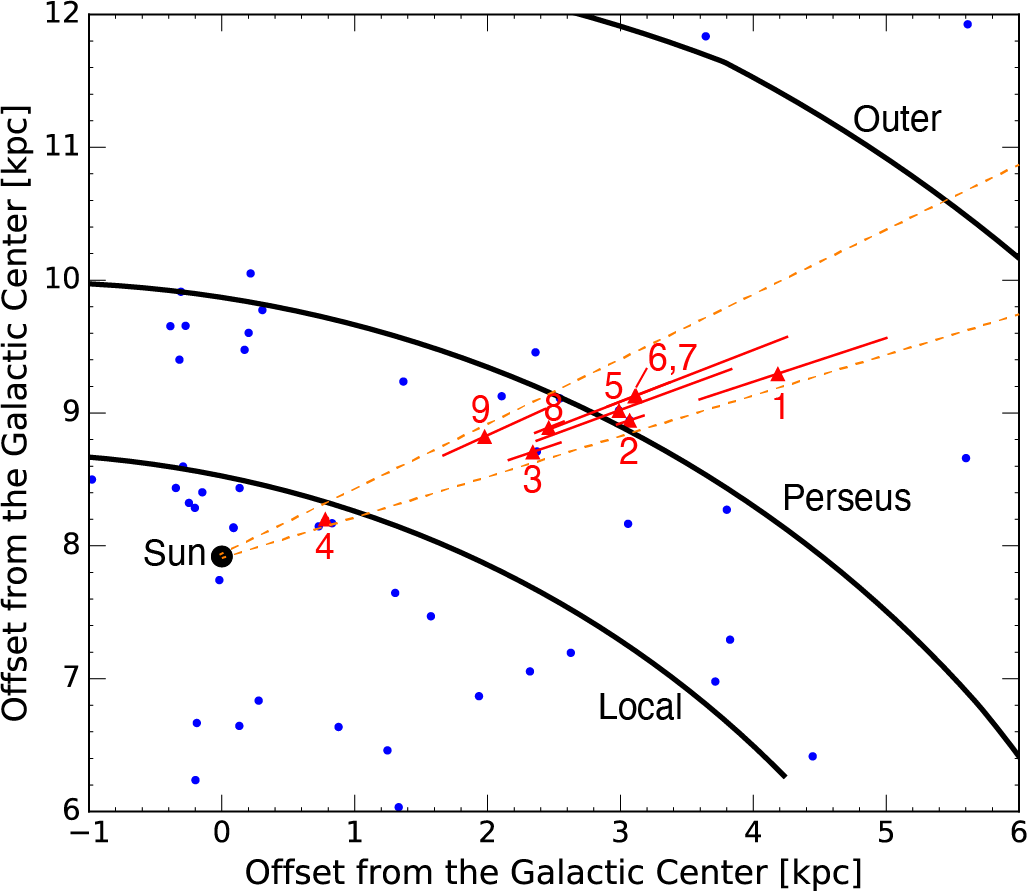}
\caption{The face-on distribution of molecular clouds {as viewed} from the north Galactic pole, based on VLBI astrometry results. The red triangles {and error bars} indicate the position and { uncertainty} of each molecular cloud analyzed in this paper. Blue dots indicate the position of maser sources presented in \cite{2020PASJ...72...50V}. The Black curves show the Local, Perseus, and Outer spiral arms \citep{2019ApJ...885..131R}. The black dot shows the position of the Sun located at 7.92 kpc from the Galactic center \citep{2020PASJ...72...50V}. The orange dotted lines present the Galactic longitude range between $l=\timeform{107D}$ and $\timeform{116D}$. Alt text: The face-on distribution of molecular clouds based on VLBI astrometry results.
\label{faceon}}
\end{figure*}

 Figure~\ref{faceon} shows the face-on distribution of each molecular cloud {containing} H$_2$O maser sources presented in this paper.
 {The Cepheus A molecular cloud in the Local arm (Group A) and NGC 7538 in the Perseus arm (Group B) have masses of $\sim 10^5\ M_{\odot}$, making them more massive than the other molecular clouds in our sample. The eight molecular clouds in Group B are distributed at different positions within the Perseus arm.} {We find a correlation between the size parameters and their relative positions with respect to the Perseus arm. However, this correlation is unlikely to reflect a physical relationship and is more likely an artifact, since the size parameters are proportional to the distance.}
 These results suggest that the physical properties of molecular clouds in the Cepheus--Cassiopeia region may not vary significantly with location within the Perseus arm.
{Based on the new parallax measurement presented in this paper, we have} revealed that IRAS 23385+6053 is located {in front of} the Perseus arm {as seen from the Sun. This object exhibits radially inward motion (i.e., $U_{\rm s} = 25\pm8$ \kms) and lags behind the Galactic rotation (i.e., $V_{\rm s} = -12 \pm 7 $ \kms), which is consistent with previous VLBI astrometry results of the Perseus arm (e.g., see \citealp{2019ApJ...876...30S}).} 

\citet{2022ApJ...925..201P} presented a face-on view of molecular clouds in the Perseus arm between $l=\timeform{135D}$ and $l=\timeform{160D}$. Their study was based on a three-dimensional dust map constructed using distance measurements from Gaia annual parallaxes, together with stellar photometry from Pan-STARRS 1 and 2MASS, as presented by \citet{2019ApJ...887...93G}. They showed that, for a series of molecular clouds located close to one another in the longitude--velocity diagram, the actual distances vary significantly, and that the molecular clouds in the Perseus arm extend over $\sim 3$ kpc in radial distance (see Figures 4 and 5 in \citealp{2022ApJ...925..201P}). 
In this study, we investigated the three-dimensional structure of molecular clouds in the Cepheus--Cassiopeia region between $l=\timeform{107D}$ and $ l = \timeform{116D}$ using distance measurements based on the annual parallaxes of H$_2$O maser sources (Figure~\ref{faceon}). {We find that molecular clouds in Group B, corresponding to the Perseus arm in the longitude-velocity diagram, are distributed within a range of approximately 2 kpc from the center of the Perseus arm, which considering the uncertainty of the trigonometric distances. }
{Analysis of CO line data in combination with maser sources, as presented in this paper, will be essential for future studies to reveal the relationship between the physical properties and the three-dimensional distribution of molecular clouds throughout the Milky Way.}

\section{Conclusions}
We {used VERA to conduct} VLBI observations to measure the trigonometric parallax of H$_2$O maser sources in the outer massive star-forming region IRAS 23385+6053{, located in} the Cepheus and Cassiopeia region. The annual parallax is $\pi=0.460 \pm 0.086$~mass, which corresponds to {a} distance of $2.17^{+0.50}_{-0.34}$ kpc{, which is} at about half the kinematic distance of 4.9 kpc reported {in} previous studies. The proper motion of IRAS 23385+6053 is ($\mu_{\alpha}\cos{\delta}$,$\mu_{\delta}$)=($-3.73\pm0.53$, $-2.0{7}\pm0.73$) mas yr$^{-1}$. We also analyzed the CO line archival data obtained by the FCRAO radio telescope. The {LSR} velocities coincide between molecular {gases} and H$_2$O maser sources in the Cepheus and Cassiopeia region. 
{Based on distances obtained from annual parallaxes measured by VERA and VLBA, the three-dimensional distribution of GMCs was revealed.} 
{Our results suggest that, taking into account distance errors, the molecular clouds in the direction of Cepheus and Cassiopeia region in the Perseus arm are distributed within a range of approximately 2 kpc from the center of the Perseus arm.}

\section*{Acknowledgments}
{We are grateful to the referee Dr. Ye Xu of Purple Mountain Observatory for carefully reading our manuscript and giving us thoughtful suggestions, which greatly improved this paper.}
We are grateful to VERA project members for their support on VLBI observations and data reduction.
We used the software packages of Astropy \citep{2013A&A...558A..33A,2018AJ....156..123A,2022ApJ...935..167A}, NumPy \citep{2011CSE....13b..22V}, Matplotlib \citep{2007CSE.....9...90H}, and APLpy \citep{2012ascl.soft08017R}.
This publication makes use of data products from the Wide-field Infrared Survey Explorer, which is a joint project of the University of California, Los Angeles, and the Jet Propulsion Laboratory/California Institute of Technology, funded by the National Aeronautics and Space Administration.
The research presented in this paper has used data from the Canadian Galactic Plane Survey, a Canadian project with international partners, supported by the Natural Sciences and Engineering Research Council.


\end{document}